\documentstyle[twoside,fleqn,espcrc2,epsf]{article}

\newcommand{\beq}{\begin{equation}}
\newcommand{\eeq}{\end{equation}}
\newcommand{\bea}{\begin{eqnarray}}
\newcommand{\eea}{\end{eqnarray}}
\newcommand{\bi}{\bibitem}

\newcommand{\NP}{NP}
\newcommand{\PL}{PL}
\newcommand{\PR}{PR}
\newcommand{\PRL}{PRL}

\title{\vspace{-4.0cm} 
\begin{flushright}
{\normalsize\tt BNL-HET-98/32}\\     
\end{flushright}
\vspace*{2.0cm}
Light quark masses using domain wall fermions
\thanks{Talk by M.W.\ at Lattice '98, Boulder, CO, USA.
Work supported in part by DE--AC02--98CH10886}
}

\author{Tom Blum\address{Department of Physics, Brookhaven
	National Laboratory, Upton, NY 11973, USA},
	Amarjit Soni$^{\rm a}$, 
	and Matthew Wingate\address{RIKEN BNL Research Center,
	Brookhaven National Laboratory, Upton, NY 11973, USA}        
}
       
\begin{document}

\begin{abstract}
We compute the one--loop self--energy correction 
to the massive domain wall quark propagator.  Combining
this calculation with simulations at several 
gauge couplings, we estimate the strange 
quark mass in the continuum limit. 
The perturbative one--loop mass renormalization is 
comparable to that for Wilson quarks and considerably smaller
than that for Kogut--Susskind quarks. Also, scaling violations
appear mild in comparison to other errors at present.
Given their good chiral behavior and these features,
domain wall quarks are attractive for evaluating the
light quark masses.  Our preliminary quenched result is
$m_s(2$ GeV) = 82(15) MeV in the ${\overline{\rm MS}}$ scheme.
\end{abstract}

\maketitle

\section{INTRODUCTION}

Computing light quark masses is a high priority.
Present lattice predictions of $m_l \equiv
(m_u+m_d)/2$ and $m_s$ give differing results depending on the
particular method.  
Domain wall (DW) fermions \cite{ref:KAPLAN,ref:SHAMIR} respect the chiral
symmetries of the continuum exactly in the limit $N_s\rightarrow\infty$,  
$N_s$ being the number of sites in a fictitious extra dimension.
This is an especially attractive feature for simulating
light quark physics where chiral symmetry is crucial 
\cite{ref:SHAMIR,ref:FURSHAM}.
Furthermore, simulations have demonstrated
that $N_s\sim 10$ for $\beta \ge 6.0$ suffices for very good
chiral behavior rendering DW fermions quite practical \cite{ref:BLUM}.
We report on first results for the strange quark mass
using DW fermions.

\section{ACTION}

We use the boundary fermion variant \cite{ref:SHAMIR} of the
DW formulation.
The fifth dimension has a finite number of sites $N_s$;
a light right--handed mode is bound to the 4--d surface 
at one end and a light left--handed mode at the other end.
A 5--d mass $M$ determines the strength of the
binding, and a parameter $m$ controls the coupling between the two
5--d boundaries.

The chiral symmetry breaking is due to the explicit term proportional to
$m$ and to implicit mixing between the two modes which should be suppressed
$\sim e^{-\alpha N_s}$, where $\alpha > 0$. 
Therefore the latter can be made arbitrarily small
compared to the former by increasing $N_s$.
For example, at tree level the quark mass is \cite{ref:SHAMIR,ref:VRANAS}
\beq
a m_q ~=~ M(2-M)\Big( m + |1-M|^{-\alpha N_s}\Big).
\label{eq:mq0}
\eeq
If, in the free theory, $M$ 
is in the range $0<M<2$, there is a single light mode fixed to either boundary.
More properties of DW fermions are discussed in \cite{ref:BLUM_LAT98}
and references therein.

\section{RENORMALIZATION}

Interactions renormalize the five--dimensional mass $M$ additively; 
thus the range $0<M<2$ is shifted by an amount $M_c$.
In perturbation theory $M_c$ can be computed from
terms proportional to $a^{-1}$.  Numerically, the tadpole graph
gives larger contributions to the self--energy than 
the half--circle graph; so one way to estimate $M_c$ is to
use the tadpole term alone, giving $M_c^{\rm tad} = 12.6\alpha_s$
\cite{ref:AOKI}.  
Noting the importance of $M_c$ and
the limitations of perturbation theory, a nonperturbative determination
of $M_c$ is more desirable.  
Through the overlap formulation \cite{ref:NNBIG} the 
Wilson--Dirac operator defines a transfer
matrix which governs propagation in the 5th dimension.  Therefore
$M_c$ can be determined directly from the critical Wilson hopping parameter
$\kappa_c^W$ \cite{ref:EHN}:
\beq
M_c^W = 4 - (2\kappa_c^W)^{-1}.
\eeq
Using data from \cite{ref:BBS} we give $M_c^W$ relevant to
this work in Table \ref{tab:mcresults}.
The superscript is to distinguish this calculation of $M_c$
from other estimates.  
For reasonable definitions of the coupling constant $\alpha_s$ (see later),
we find $M_c^{\rm tad} \approx M_c^W$.

\begin{figure}
\vspace{2.0in}
\includegraphics{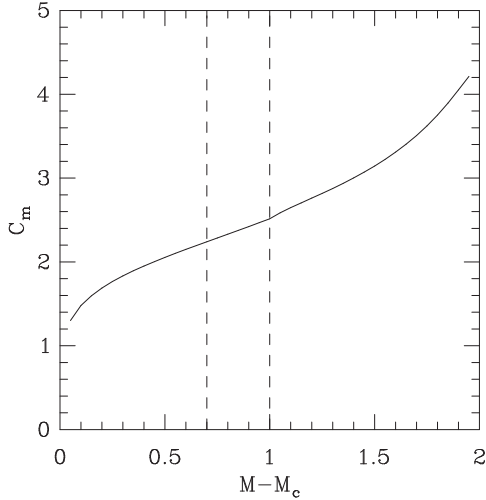}
\caption{ The matching coefficient $C_m$.
The dashed lines indicate the range of $M-M_c$ where
simulations have been performed. }
\label{fig:c_m}
\end{figure}

In order to compute the quark mass renormalization,
we have extended the explicit one--loop calculation \cite{ref:AOKI}
of the DW fermion self--energy to nonzero fermion mass
\cite{ref:SOON}.  The renormalization factor
$Z_m$ can be defined by equating the one--loop fermion propagators
on the lattice and the continuum \cite{ref:GONZ}.
The result is
\beq
Z_m = 1 - (2\alpha_s/\pi)\Big[ \ln(\mu a) - C_m \Big],
\label{eq:Zm}
\eeq
where the DW quark masses are computed with a lattice spacing $a$
and the $\overline{\rm MS}$ quark masses at a scale $\mu$.
$C_m$ depends only on $M-M_c$ and is plotted
in Fig.~\ref{fig:c_m}.  The $C_m$ for the $M-M_c^W$ used in our
simulations are shown in Table~\ref{tab:mcresults} and should
be compared to 2.16, 3.22, and 6.54 for Wilson, 
Sheikholeslami--Wohlert (SW), and Kogut--Susskind (KS) fermions.

\begin{table}
\caption{\label{tab:mcresults} Summary of simulation parameters and
results.
$m_s^{\rm LAT}$ and $m_s^{\overline{\rm MS}}(2$ GeV)
are in MeV. }
\begin{center}
\begin{tabular}{l|ccc}
& $\beta = 6.3$ & $\beta=6.0$ & $\beta=5.85$ \\ \hline
\# configs. & 18 & 33 & 18 \\
volume & $24^3\!\times\! 60$ & $16^3\!\times\! 32$ & $16^3\!\times\! 32$ \\
$N_s$  & 10 & 10 & 14 \\
$M$ & 1.5 & 1.7 & 1.7 \\
$M_c^W$ & 0.708 & 0.819 & 0.908 \\
$C_m$ & 2.35 & 2.47 & 2.42 \\
$a^{-1}$ (GeV) & 3.4(3) & 2.39(12) & 1.69(10) \\
$m|_{\rm strange}$ & 0.020(3) & 0.027(3) & 0.041(6) \\
$m_s^{\rm LAT}$ & 64(11) & 62(8) & 66(10) \\
$\langle{\rm Tr}~U_{\rm plaq}/3\rangle$ & 0.6224 & 0.5937 & 0.5751 \\
$\alpha_{\overline{\rm MS}}(2/a)$ & 0.131 & 0.146 & 0.157 \\
$m_s^{\overline{\rm MS}}(2$ GeV) & 85(15) & 81(10) & 83(13)
\end{tabular}
\end{center}
\end{table}

\section{QUARK MASS}

\begin{figure}
\vspace{2.0in}
\includegraphics{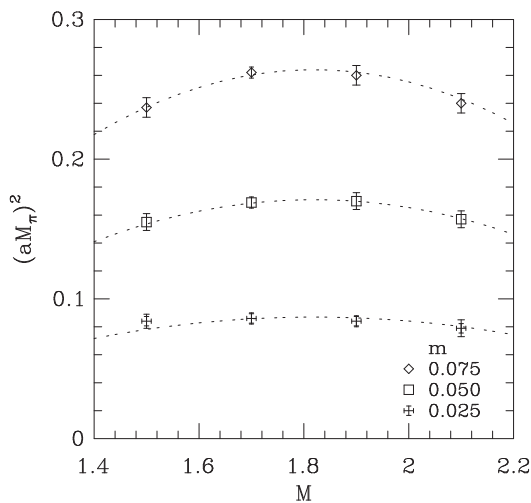}
\caption{ Pion mass squared as a function of $M$ at
$\beta = 6.0$.  $N_s=14$ except for the $M=1.7$, $m=0.075$ point
where $N_s=10$.  The dotted lines are fits to
Eqn.\ \ref{eq:mpi2fit}. }
\label{fig:mpi2_m0}
\end{figure}

We have computed the pion mass and decay constant on
a few dozen quenched configurations at $\beta = 5.85$, 6.0, and 6.3.
We give other simulation parameters in Table~\ref{tab:mcresults}.
In addition we studied the pion mass as a function of both $m$ and
$M$ at $\beta=6.0$.  
For each $M$, the pion mass squared extrapolates linearly to zero
at $m=0$ within errors as in \cite{ref:BLUM}.  
Fig.~\ref{fig:mpi2_m0} shows the dependence of the pion mass squared
on $M$.  In lowest order chiral perturbation theory 
$M_\pi^2 \sim m_q$.  Therefore, from Eqn.~(\ref{eq:mq0}) 
with $M\rightarrow M-M_c^W$ the pion mass squared should obey
\beq 
(aM_\pi)^2 = {\rm (const.)}(M - M_c^W)(2 - M + M_c^W).
\label{eq:mpi2fit}
\eeq
The dotted lines are fits to (\ref{eq:mpi2fit}) and have 
good $\chi^2$'s.  At present the data do not permit a more
general fit.

To numerically compute the renormalization constant (\ref{eq:Zm}),
one must choose a definition for the coupling constant $\alpha_s$ and the
scale $\mu$ at which it is evaluated.  We follow~\cite{ref:LANL}
and use $\alpha_{\overline{\rm MS}}(\mu)$.  Specifically,
we compute $\alpha_V(3.41/a)$ from the plaquette and convert to
$\alpha_{\overline{\rm MS}}$ perturbatively \cite{ref:LM}.
 We run that coupling
constant to $\mu$ to apply (\ref{eq:Zm}) and finally run
$m_{\overline{\rm MS}}$ perturbatively to 2 GeV.
We choose $\mu = 2/a$ for the matching; the final result varies
by 2 MeV for $0.5 < \mu a < \pi$.
We use the physical pion decay constant to set the lattice spacing
and the physical kaon mass to fix the parameter $m$ to the strange
quark system.

\begin{figure}
\vspace{2.0in}
\includegraphics{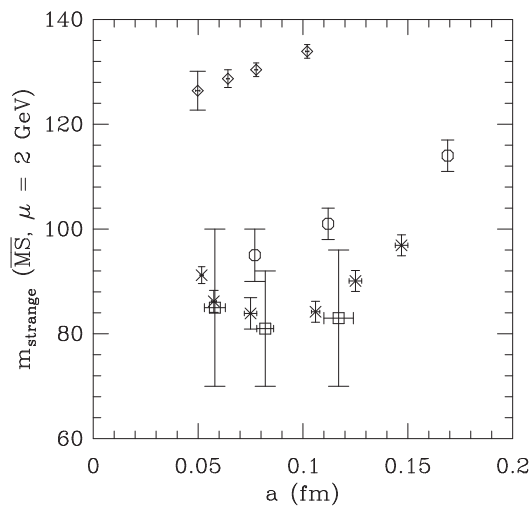}
\caption{ Strange quark mass in $\overline{\rm MS}$ scheme at 2 GeV. 
Our results are displayed
as squares, Wilson fermions as diamonds~\cite{ref:CPPACS_LATEST},
SW fermions as circles~\cite{ref:FNAL},
and KS results as asterisks~\cite{ref:JLQCD_STAG,ref:KIM_OHTA}.}
\label{fig:mstrange_msbar}
\end{figure}

\setcounter{footnote}{0}
In Fig.~\ref{fig:mstrange_msbar} we plot our final results for
the strange quark mass along with recent results using other
actions.  
The results for DW fermions appear to be in rough agreement
with other lattice calculations of the strange quark mass
using the kaon mass and the perturbative 
$Z_m$.\footnote{For other recent results see \cite{ref:KENWAY}.}
Within our rather large errors, 
the DW results appear scale independent, so for now
we take a weighted average.
Our statistical errors are large
due to our small data set. We attribute a 2 MeV error
to the ambiguity in $q^*$ and tentatively assign a 15\% error
due to determining $a^{-1}$ from $f_\pi$ \cite{ref:BLUM}.
Quenching errors are not included at present.
We add the systematic and statistical errors in quadrature
to obtain the preliminary result for the 
$\overline{\rm MS}$ quark mass: $m_s(2~{\rm GeV}) = 82(15)$ MeV.

We find that DW fermions are a viable method 
of studying the light quark masses.  We have
shown that perturbative corrections to the mass renormalization
are under control, the lattice quark mass depends on $m$ and $M$
as expected, and scaling violations appear mild at this
stage.

We thank T.\ Bhattacharya for discussions.

\end{document}